\documentstyle[11pt,paspconf,psfig]{article}
\begin{document}

\def\ol{\Omega_{\Lambda}}
\def\ob{\Omega_{b}}
\def\obh{\Omega_{b}h^{2}}
\def\oc{\Omega_{CDM}}
\def\om{\Omega_{m}}
\def\t{t_{o}}
\def\gtrsim{\mathrel{\hbox{\rlap{\hbox{\lower4pt\hbox{$\sim$}}}\hbox{$>$}}}}
\def\lesssim{\mathrel{\hbox{\rlap{\hbox{\lower4pt\hbox{$\sim$}}}\hbox{$<$}}}}
\def\lsim   {\wisk{_<\atop^{\sim}}}
\def\gsim   {\wisk{_>\atop^{\sim}}}
\def\etal{{\em et al.~}}
\def\apj{{\em Ap.J.}}
\def\apjl{{\em Ap.J.L.}}
\def\mnras{{\em M.N.R.A.S.}}
\def\aa{{\em A \& A}}
\def\be{\begin{equation}}
\def\ee{\end{equation}}

\title{CMBology}

\author{Charles H. Lineweaver}
\affil{School of Physics, University of New South Wales\\
    Sydney, NSW, 2052 Australia}

\begin{abstract}
The details of the CMB power spectrum are being revealed through the
combined efforts of the world's CMBologists. The current data set 
constrains several cosmological parameters. A combination with other 
(non-lensing) constraints yields estimates of the cosmological constant: 
$\ol = 0.65 \pm 0.13$, the mass density: $\om = 0.23 \pm 0.08$ and the 
age of the Universe: $\t = 13.4 \pm 1.6$. Lensing data is not yet comfortable 
with these values.
\end{abstract}



\section{Seeing Sounds in the CMB (see Figure 1)}

As the Universe cools it goes from being radiation dominated to matter 
dominated. The boundary is labelled `$z_{eq}$' on the right side of Fig.~1.
As the Universe cools further, electrons and protons combine, 
thereby decoupling from photons during the redshift interval `$\Delta z_{dec}$'. 
The opaque universe 
becomes transparent. Present observers, on the left of Fig.~1, 
look back and see hot and cold spots on the surface of last scattering.
But where did the hot and cold spots come from?

At $z_{eq}$, dark matter over-densities begin to collapse gravitationally.
The photon-baryon fluid (grey) falls (inward pointing arrows) into the dark 
matter potential wells -- gets compressed (dark grey) and then rebounds (outward 
pointing arrows) due to photon pressure support -- leaving less dense regions 
(white) at the bottoms of the wells, then recollapses and so on. In the 
observable interval $\Delta z_{dec}$, the phases at which we see these 
oscillations depend on their physical size. Four different sizes with four 
different phases are shown in Fig. 1. From top to bottom: maximum Doppler 
inward velocities, maximum adiabatic compression, maximum Doppler outward 
velocities, maximum adiabatic rarefaction. The corresponding power spectrum 
of the CMB, $C_{\ell}$, is shown on the left. Notice that the peaks in the total 
power spectrum are due to adiabatic compression and rarefaction, while the valleys 
are filled in by the relatively smaller Doppler peaks. Although we have used the 
example of dark matter over-densities, we are in the regime of small amplitude linear 
fluctuations and so  dark matter under-densities produce the same power spectrum, 
i.e., the first and largest peak in the total spectrum is produced by equal numbers 
of hot and cold spots on the surface of last scattering.  When we see hot and cold 
spots in the CMB we are seeing sound: acoustic adiabatic compressions and 
rarefactions, visible across 13 billion years of vacuum. 

\clearpage
\begin{figure}[b]
\centerline{\psfig{figure=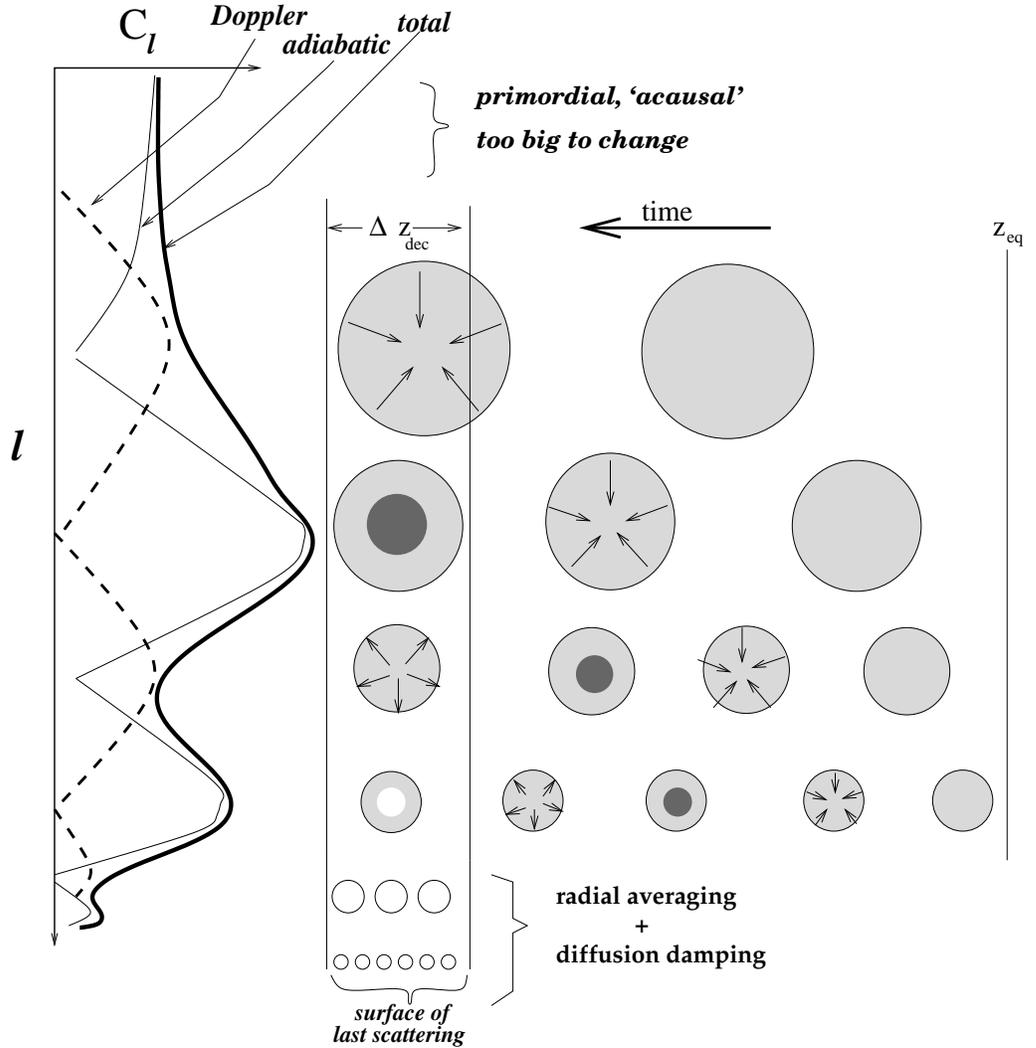,angle=0,height=15cm,width=13.5cm,rheight=14cm,rwidth=13.5cm}}
%
\caption{Seeing sound. The observer is on the left.
Sound waves in the photon-baryon fluid create bumps in the 
CMB power spectrum.
The grey spots are cold dark matter potential wells which
initiate infall and then oscillation of the photon-baryon fluid 
in these wells.
The Doppler and adiabatic effects make the 
sound visible in the radiation when the baryons decouple from the 
photons during the interval marked $\Delta z_{dec}$.
These bumps are analogous to the standing waves of the resonant frequencies
of a good shower or of a plucked string and may be the oldest music in 
the Universe.
See Hu {\it etal} (1996), Tegmark (1996) and Lineweaver (1997) for details.
 } \label{fig-1}
\end{figure}

\clearpage
\section{The CMB Data (see Figure 2)}

Since the COBE-DMR discovery of hot and cold spots in the CMB (Smoot \etal 1992), 
anisotropy detections have been reported 
by more than a dozen groups with various instruments, at various frequencies and 
in various patches and swathes of the microwave sky. 
The top panel of Fig. 2 is a compilation 
of recent measurements. The COBE-DMR points (on the left) are at large 
angular scales while most recent points are trying to constrain the 
position and amplitude of the dominant first peak at $\sim 0.5$ degrees.

The three curves are: $(\Omega_{m},\Omega_{\Lambda}) = (1.0,0.0),(0.3,0.0)$ 
and $(0.3,0.7)$ (SCDM, OCDM and $\Lambda$CDM, respectively).
The $\Lambda$CDM model fits the position and amplitude of the dominant
first peak quite well. 
The largest feature in the data which doesn't fit this model 
is the low values in the $20 < \ell < 100$ region. 
Blame the PythV points.
%
The SCDM model has a peak amplitude much too low. Lowering
$H$ to $50$ would bring the peak down a further 10\% to 20\%. 
The OCDM model has the peak at angular scales too small to fit the data 
and is strongly excluded by a fuller analysis (cf. Fig. 3A but 
see Ratra \etal (1999) for a dissenting view).

The positions and amplitudes of the features in model $C_{\ell}$'s, (Zaldarriaga \& Seljak 1996), 
depend on cosmological parameters. This dependence allows measurements of the CMB power 
spectrum to constrain the parameters.  
The results I report here are based on such an analysis and may be
compared with Tegmark (1999), Efstathiou \etal (1999) and Bond \& Jaffe (1999).

A major concern of CMB measurements is galactic foreground contamination
(see the monograph ``Microwave Foregrounds'', de Oliveira-Costa \& Tegmark 1999
for a review and update).
Another concern is the analysis methods used to convert the measurements
into constraints on parameters. 
The data points in the top panel of Fig. 2 are almost independent.
A few are looking at overlapping patches of sky in similar $\ell$-bands, thus 
cosmic variance components of the error bars are correlated. Observers use similar 
instrumentation which can produce correlated systematic errors. Some calibrate 
on the same sources and therefore may share the same systematic calibration error.
Although ``there is no systematic way to handle systematic errors'' (to quote Paul Richards),
we do know that partially correlated error bars reduce the effective number
of degrees of freedom. However, the $\chi^{2}/DOF$ of the best fit is marginally too
good. Thus there is some room to reduce the $DOF$ and still find  good-fitting best-fits.

Another concern is the assumptions we make.
CMB results are subject to several well-motivated assumptions; we assume
gaussian adiabatic initial conditions. This means that when we input initial gaussian
fluctuations of the density field, we put hot baryon/photon fluid in potential wells and cold
baryon/photon fluid on potential hills, as if the compression/rarefaction mechanism discussed
in Section 1 had already been active on all scales, including super-horizon scales.
An alternative, isocurvature initial conditions, seems to be disfavored by the data.
We also assume that the topological defect mechanisms have not played an important role.
These and other assumptions are beginning to be looked at more carefully.

\clearpage
\begin{figure}
\vspace{-0.5cm}
\centerline{\psfig{figure=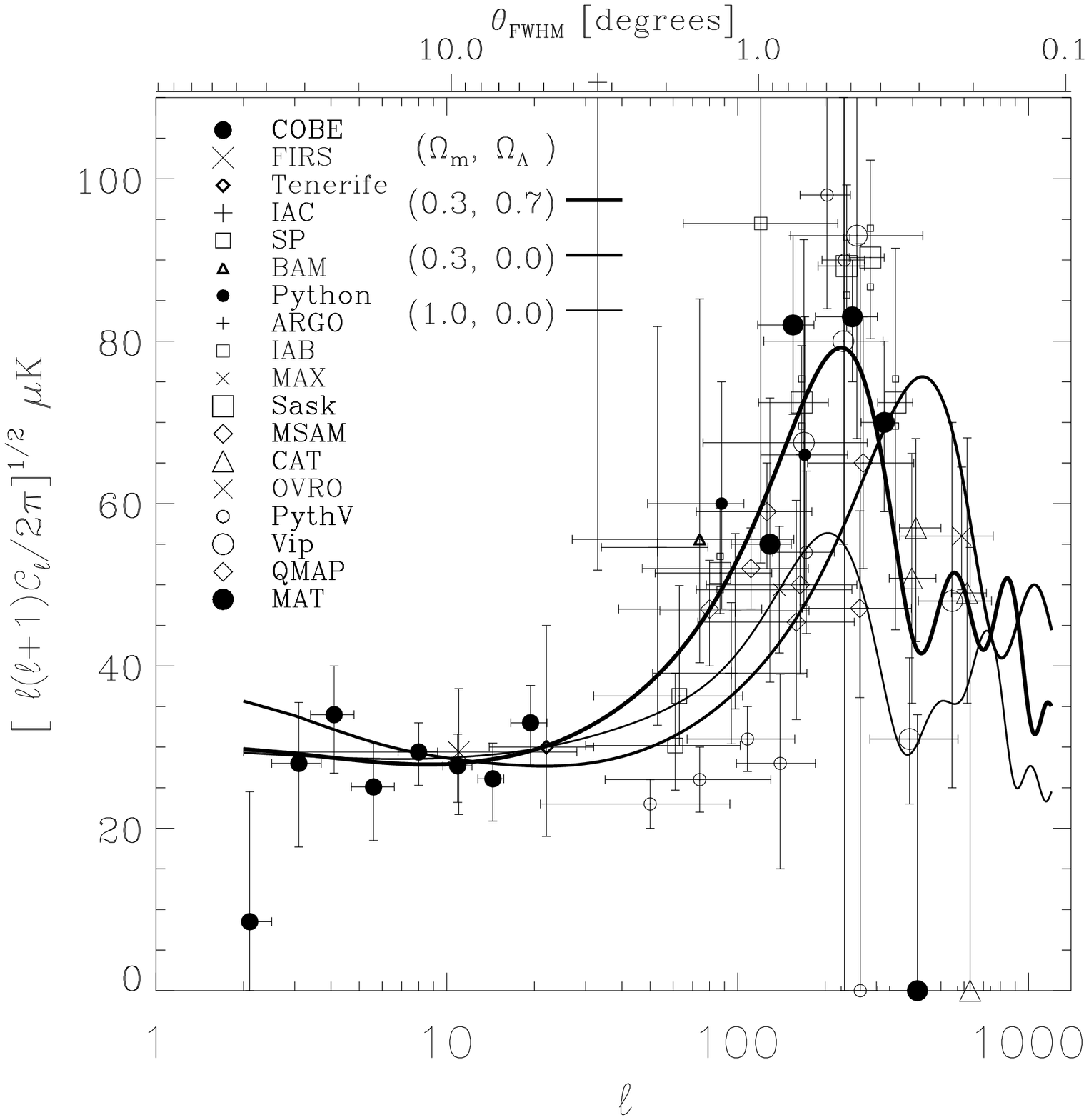,angle=0,height=9cm,width=11cm}}
\centerline{\psfig{figure=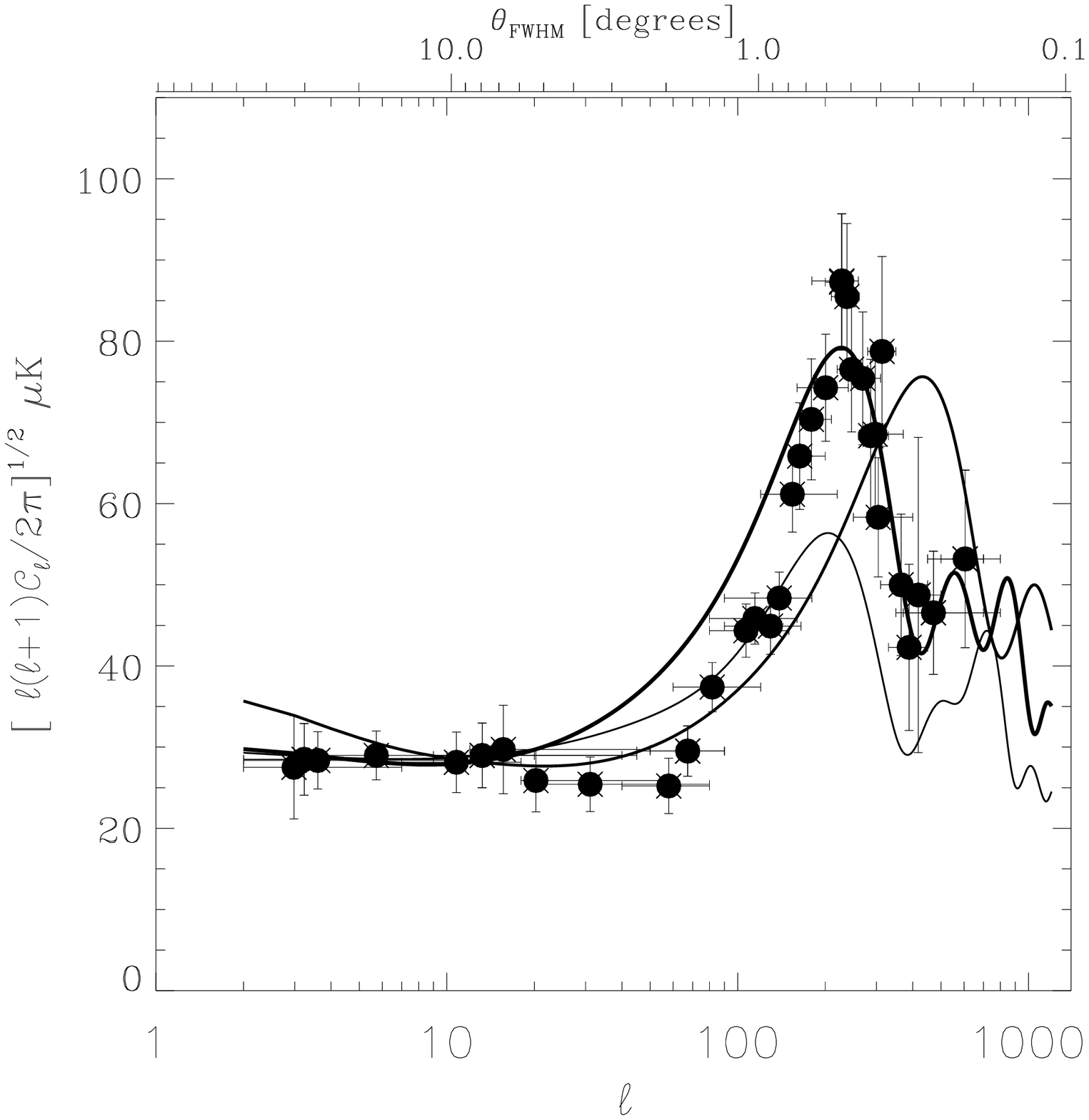,angle=0,height=8.5cm,width=10.5cm}}
\vspace{-0.7cm}
\caption{{\bf Top:}  Measurements of the CMB power spectrum. 
New sets of points are being added to the top panel every month or so. 
The three curves represent the three most popular models and are the same in both
panels. All three models have $h = 0.65$ and are normalized 
to agree with the large angular scale measurements.
{\bf Bottom:}
To reduce some of the scatter and to obtain a model-independent
curve preferred by the data, I have averaged the points in the top panel
into 8 bins and then rebinned 5 times.  Just connect the dots. 
The 40 points in the lower panel are, therefore, not independent, so do not attempt a $\chi$-by-eye.} 
\label{fig-2}
\end{figure}

\clearpage

\section{Cosmological Parameters from the CMB, SN,\\
 Galaxy Clusters $\ldots$  and Even Lenses (See Figure 3)}

Fig. 3  contains multiple constraints in the $\om - \ol$ plane and is difficult to read.
I took Fig. 3 of Lineweaver (1999) and overlayed all the lensing constraints
I could find in the literature. In A, the elongated triangle (upper left to lower right)
is the $68\%$ confidence level (CL) region  preferred by the CMB data (the 95\% and 99\% CL regions
are also shown). The dark grey triangle in the lower right is the $68\%$ CL  from lensing
reported in Falco \etal (1998). The region extends to the x axis but is hidden behind the CMB 68\% CL.
The light grey region (upper right to lower left) is the $95\%$ CL from Falco \etal (1998).
The small elongated almond shape in F is the $68\%$ CL region based on the CMB and 4 other sets of
observations (B,C,D \& E).  When this region is projected onto the $\om$ and
$\ol$ axes it yields 
\begin{itemize}
\item $\ol = 0.65 \pm 0.13$
\item $\om = 0.23 \pm 0.08$ 
\end{itemize}
These results do not include lensing constraints because there seems to be some disagreement
about what the lensing constraints are.
Notice that in A, the CMB and lensing constraints overlap in the region around $(\om,\ol) = (1.0,0.0)$.
However in this region the age of the Universe is $\sim 10$ Gyr (too young) and this age problem can only
be remedied by making the Hubble constant $\sim 50$
(too low). The dark and light grey regions in B through E are other published lensing
results: B: Chiba \& Yoshii 1999 and Cheng \& Krauss 1998,
C: Quast \& Helbig 1999 and Helbig \etal 1999, D: Cooray (2000),
E: Cooray (1999a,b). 
In general the lensing constraints are parallel to the lines of constant age
(and also to the SN constraints in B). Only D gives a lensing result with little $\ol$
dependence.

To a large extent these lensing constraints depend on the same or similar
data so they are not independent. The constraints in A, C and E are very similar; standard CDM 
is favored. The lensing constraints in B and D however are in very good agreement with the
results shown in F.
Kochanek \etal (1999) have  criticized the panel B results based largely on
different estimates of the Schecter function parameters $\alpha$ and $B_{*}$.

Suppose that $\Lambda$CDM is the correct cosmology, i.e., that the combination of data that 
produced the contours in F are correct.
What mistake has been made in interpreting the lensing data, producing the constraints in A, C and E?
Kochanek seems to think that people are barking up the wrong tree: lensing models. They should be 
barking up the galactic evolution tree, i.e., that our understanding of the lensed population is more important
than the degree of central concentration or other aspects of the lens models.

\begin{figure}
\centerline{\psfig{figure=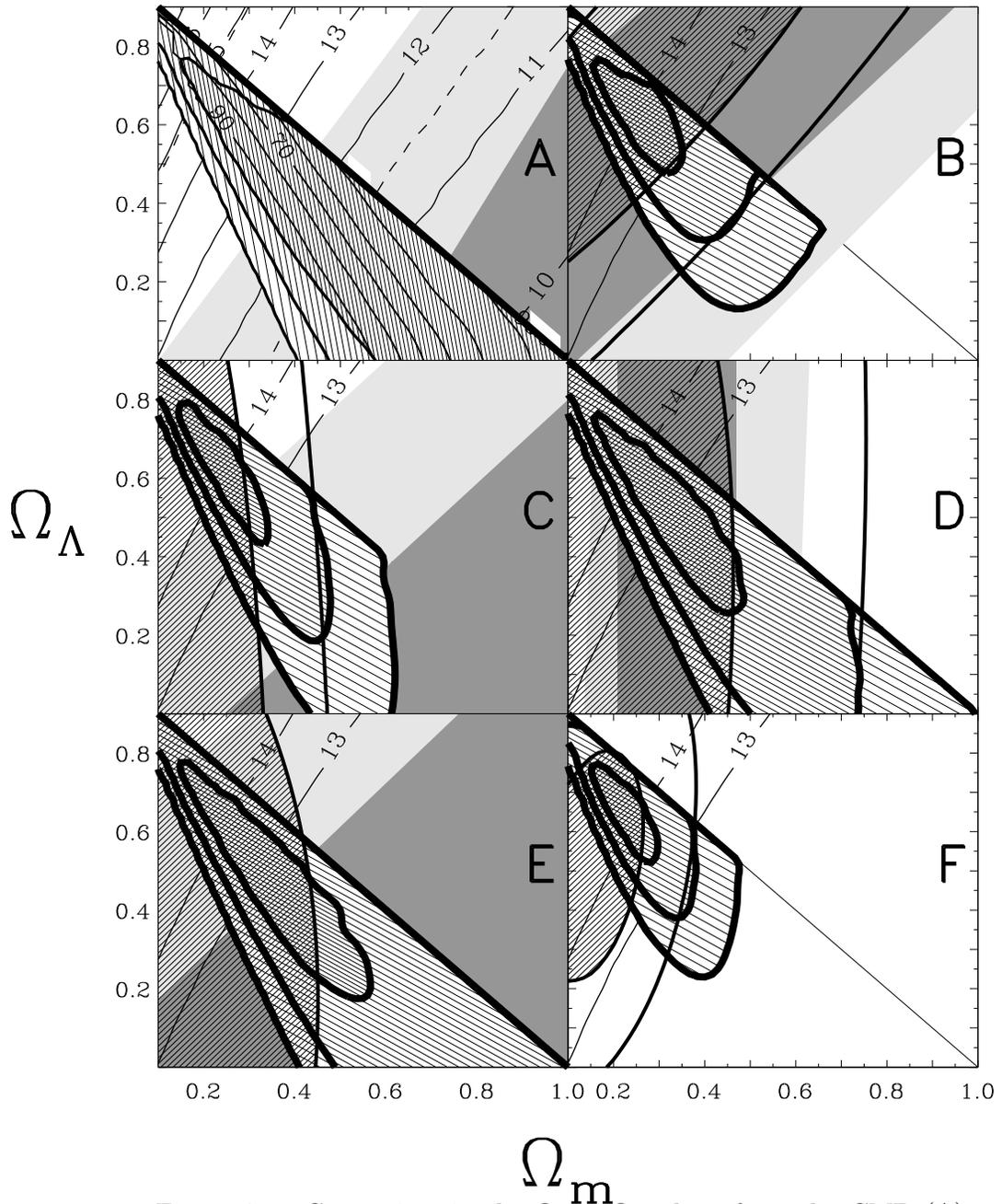,angle=0,height=17cm,width=16cm,rheight=18.1cm,rwidth=17.5cm}}
\vspace{-1cm}
\caption{Constraints in the $\Omega_{m} - \Omega_{\Lambda}$ plane from the CMB 
(A), supernovae (B), cluster M/L (C), cluster evolution (D), double lobed radio 
sources (E) and all combined (F). The thickest contours are from combining these constraints
with the CMB constraint in A.
The results cited in the abstract: 
$\ol = 0.65 \pm 0.13$ and $\om = 0.23 \pm 0.08$ come from F.
The contours labeled `10' through `14' (Gyr) are the iso-age contours for $h = 0.68$; the 13 and 
14 Gyr contours are repeated in all panels.
The contours within the CMB 68\% CL are the best-fitting $H$ values.
See Lineweaver (1999) for details of these constraints. 
} \label{fig-3}
\end{figure}

I came to this conference hoping to hear what the latest new $\om - \ol$ constraints from
lenses were. It seems that people are trying hard to increase the sample size and conscientiously
wrestling with systematics of the assumptions required to convert lensing number counts into 
constraints on $\om$ and $\ol$. 
I was hoping to hear some heated discussion about the different lensing constraints in the $\om - \ol$ plane,
but the authors of the lensing constraints in B were not at this conference to defend themselves
against the collective derision of rival data analysts.
We can all agree however that we need more lenses. The JVAS/CLASS survey and the CASTLES follow-up survey seem 
to be doing just that.


%
%
%
%

%
%
\subsection{The Age of the Universe (See Figures 4 \& 5)}

The age of the Universe can be determined from General Relativity via the Friedmann equation.
Estimates for three parameters are needed: the mass density, the cosmological constant and Hubble's
constant; that is, $\t= f(\om, \ol, h)$. 
The focus of Lineweaver (1999) was to jointly constrain $\om$, $\ol$ and $h$ 
using the CMB and 6 other independent data sets.
The result, $\t = 13.4 \pm 1.6$ Gyr, is shown in Fig. 4 and can be compared to
the ages of various models in Fig. 5.

\begin{figure}
\vspace{-1cm}
\centerline{\psfig{figure=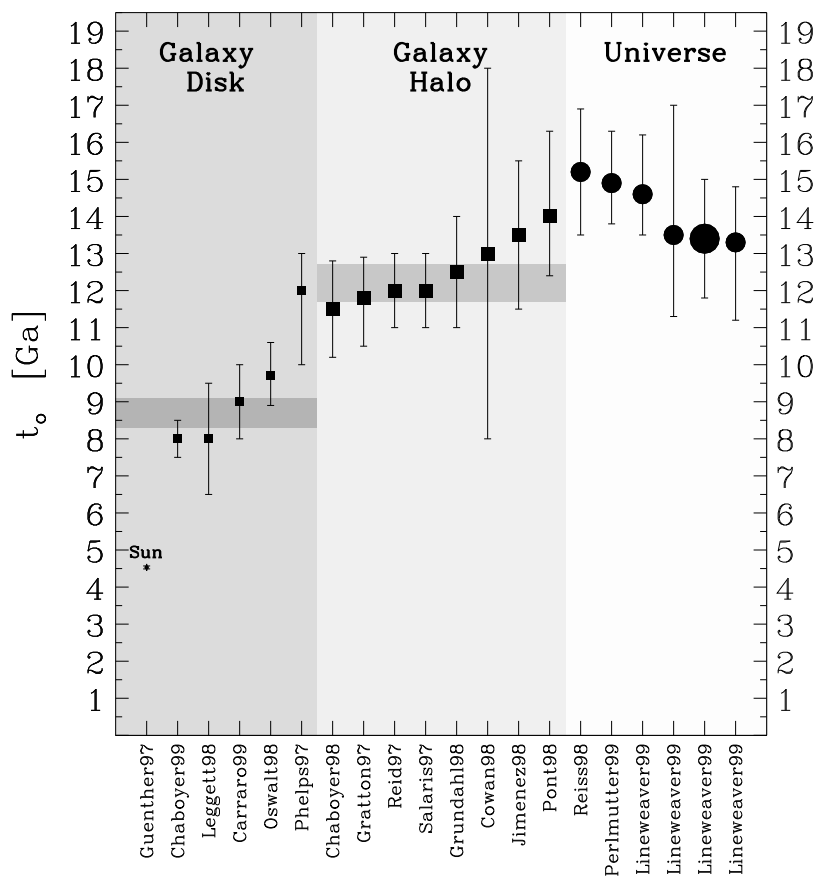,angle=0,height=12cm,width=12cm}}
\vspace{-0.3cm}
\caption{ A compilation of recent measurements of the age of the disk of the Milky Way, the halo of the
Milky Way and of the Universe. Estimates of the age of the Universe are based on
estimates of $\om$, $\ol$ and $h$. Galactic age estimates  
are direct in the sense that they do not depend on cosmology. Horizontal grey bands are averages.
The largest dot at $13.4 \pm 1.6$ Gyr is the main result of the Lineweaver (1999) paper.
It is shaded grey on the x-axis of the next figure.} \label{fig-4}
\end{figure}

\clearpage
\begin{figure}
\centerline{\psfig{figure=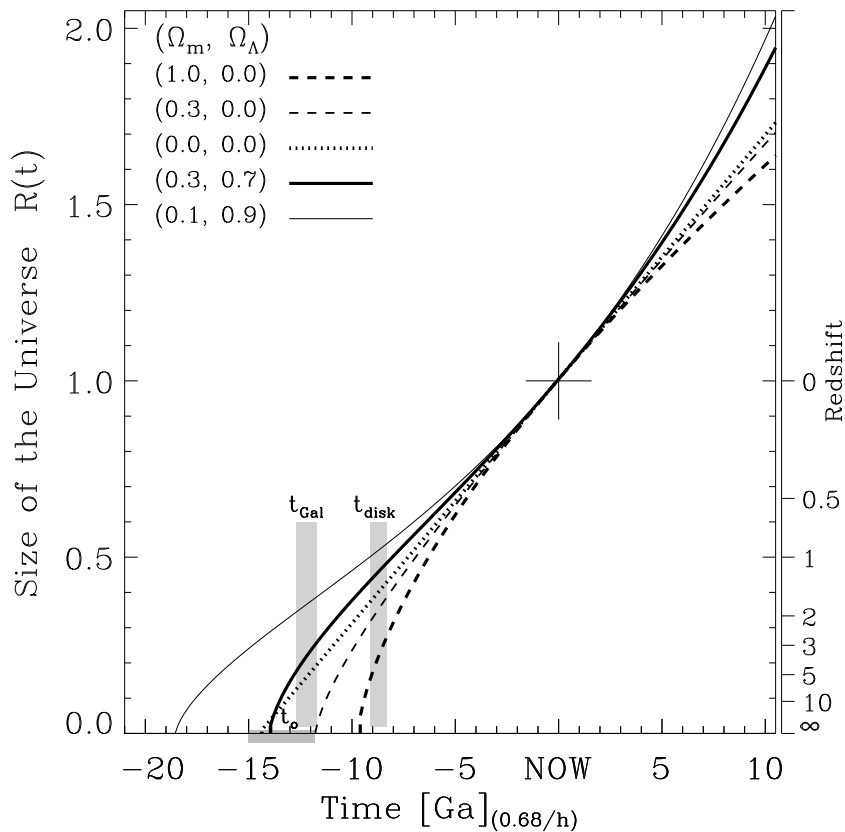,angle=0,height=12cm,width=12cm,rheight=12.0cm,rwidth=12cm}}
\vspace{-1.cm}
\caption{The ages of model universes. Any reasonable universe should be older than
the halo of our Milky Way. The critical density universes $(\om, \ol) = (1.0,0.0)$ favored by some
lensing analysts 
have severe age problems unless $h < 0.55$.
$h = 0.68$ has been used here. 
} \label{fig-5}
\end{figure}



\acknowledgments

I acknowledge a Vice-Chancellor's Research Fellowship at the University of
New South Wales. I thank Tereasa Brainerd for coherently lensing my geodesic into Boston
for a well-organized friendly conference.

\begin{question}{Prof\ Turner}
We've been seeing likelihood contour plots like yours for years -- they always 
change when new data comes out and represent some kind of cosmology du jour, 
but not much more. 
\end{question}
\begin{answer}{Dr  Lineweaver}
Maybe you can point me to the references offline.
\end{answer}
\begin{question}{Prof \ Turner}
In any case, it's not wise to combine inhomogeneous data sets.
Bad, proemial data compromises the results.
You should only use the best data. For example, did you use Paul's low H value?
\end{question}
\begin{answer}{Dr Lineweaver}
I have used $68 \pm 10$ to represent our best estimate of H. I have been
tempted to trust my better judgement and throw out data I don't like but I've 
resisted. 
\end{answer}
\begin{question}{Dr Bridle}
You seem to have selectively chosen your data sets. For example, you have not 
included constraints from Eke's work on cluster evolution.  
\end{question}
\begin{answer}{Dr Lineweaver}
I wanted to include his work but he didn't publish it in terms of constraints on
$\om$ and $\ol$ which I could easily combine with the likelihoods I presented.
His work would, I think, broaden the cluster evolution constraints in Fig. 3D
out to the right.
\end{answer}
\begin{question}{Dr Bridle}
We have done a similar analysis of the CMB data but we have done a full
integration to properly marginalize over the nuisance parameters.
\end{question}
\begin{answer}{Dr Lineweaver}
I have not marginalized by integrating. I have followed the peaks 
in likelihood space. A good 1-D analogy which represents our different techniques  
is that you are using the mean to represent the distribution while I use the mode. 
One can argue about which is better or more robust in this context. I favor the mode 
because it allows me to include more parameters which I don't have to condition on,
$\Omega_{b}h^{2}$ for example. 
\end{answer}
\begin{question}{Prof Schecter}
In the top panel of Figure 2, is one of those curves the best-fit?
\end{question}
\begin{answer}{Dr Lineweaver}
No, I just wanted to show the three most popular models in a simple way. 
I didn't minimize the $\chi^{2}$ values for each model with respect to the other parameters.
For example, all three have $h = 0.65$, $\Omega_{b}h^{2} = 0.025$ and $Q = 18\; \mu$K.
\end{answer}
\begin{question}{polite applause}
\end{question}

%
%



\begin{references}

\reference Bond, J.R. \& Jaffe, A. 1999, astro-ph/9809043
\reference Cheng, Y.N. \& Krauss L.M.  1998, astro-ph/9810393
\reference Chiba,M. \& Yoshii, Y. 1999, astro-ph/9808321  (lensed QSO's)
\reference Cooray, A., Quashnock, J.M. \& Miller, M.C. 1999a, astro-ph/9806080 \& 9811115   (lenses in HDF)
\reference Cooray, A. 1999b, A\&A submited, astro-ph/9811448   CLASS +VLA
\reference Cooray, A. astro-ph/9904245 2000, \apj, 999, L1 (optical arcs around clusters)
\reference de Oliveira-Costa,A. \& Tegmark, M. 1999, ``Microwave Foregrounds'', A.S.P. Conf. Ser. Vol 181
\reference Efstathiou, G. \etal, 1999, MNRAS, 303, 47, astro-ph/9812226 
\reference Falco, E., Kochanek, C. \& Munoz, J.A. 1998, \apj, 494 (radio galaxies and QSO's, JVAS survey)
\reference Helbig, P. \etal 1999, A\&A Suppl. Ser. 136, 297, astro-ph/9904175
\reference Hu, W., Sugiyama, N. \& Silk, J. 1996, Nature, 386, 37
\reference Kochanek, C. \etal 1999, astro-ph/981111
\reference Lineweaver, C.H. 1997, A.S.P. Conf. Ser. Vol. 126,  `From Quantum Fluctuations to Cosmological Structures', edt. D. Valls-Gabaud \etal p 185
\reference Lineweaver, C.H. 1998, \apjlett, 505, L69
\reference Lineweaver, C.H. 1999, Science, 284, 1503-1507
\reference Navarro, J.F. \& Steinmetz, M. 1999, \apj, in press, astro-ph/9908114
\reference Quast, R. \& Helbig, P. 1999, \aa 344, 721
\reference Ratra, B. \etal 1999, \apj, 517, 549
\reference Seljak, U. \& Zaldarriaga, M. 1996, \apj, 4469, 437
\reference Smoot, G.F. \etal, 1992, \apj, 396, L1
\reference Tegmark, M. 1996, Varena, astro-ph/9511148
\reference Tegmark, M. 1999, \apj,  514, L69
\end{references}
\end{document}